\documentclass[10pt,aps,prd,twocolumn, floatfix,nofootinbib,showpacs]{revtex4}
\usepackage{graphicx}
\usepackage{color}
\begin{document}



\title{Explicit CP violation in the Dine-Seiberg-Thomas model}

\renewcommand{\thefootnote}{\fnsymbol{footnote}}

\author{
S. W. Ham$^{(1,2)}$, Seong-A Shim$^{(3)}$, and S. K. Oh$^{(4)}$
 }
\affiliation{(1) Department of Physics, Korea University, Seoul 136-701 \\
(2) School of Physics, KIAS, Seoul 130-722, Korea \\
(3) Department of Mathematics, Sungshin Women's University, Seoul 136-742 \\
(4) Department of Physics, Konkuk University, Seoul 143-701 }

\renewcommand{\thefootnote}{\arabic{footnote}}

\begin{abstract}
The possibility of explicit CP violation is studied in a
supersymmetric model proposed by Dine, Seiberg, and Thomas, with two effective dimension-five operators.
The explicit CP violation may be triggered by complex phases in the coefficients for the dimension-five operators
in the Higgs potential, and by a complex phase in the scalar top quark masses.
Although the scenario of explicit CP violation is found to be inconsistent with the experimental data at LEP2 at the tree level,
it may be possible at the one-loop level.
For a reasonable parameter space, the masses of the neutral Higgs bosons and their couplings to a pair of $Z$ bosons are
consistent with the LEP2 data, at the one-loop level.
\end{abstract}


\maketitle

\section{Introduction}

It is reasonable to assume that any phenomenological model should accommodate the violation of the CP symmetry
as one of key features, since the CP violation has been observed  in the neutral kaon system more
than four decades ago [1]
and it is one of the Sakharov conditions for the baryogenesis in cosmology to
explain the baryon asymmetry of the universe [2].
The Standard Model (SM) may explain the small CP violation in the weak interactions in terms of a complex phase in
the Cabibbo-Kobayashi-Maskawa matrix for the charged weak current [3].
However, the size of the complex phase in the SM is too small to satisfy the Sakharov conditions
with respect to the baryon asymmetry.

As an alternative source of CP violation, Weinberg has noticed that if a model possesses at least two Higgs doublets,
CP violation may occur in its Higgs sector through the mixing between the CP even and the CP odd states [4].
It is clear that the supersymmetric standard models satisfy the requirement that Weinberg has demanded,
because they have at least two Higgs doublets in order to generate the masses for
up-like quarks and down-like quarks independently [5-8].

Thus, a large number of articles have been devoted to investigate the possibility of CP violation
in supersymmetric standard models.
The minimal supersymmetric standard model (MSSM), the simplest version of supersymmetric
standard models, has just two Higgs doublets.
Therefore, in principle, the MSSM may accommodate CP violation by means of complex
phases in its neutral Higgs sector.
In practice, it has been found that CP violation is impossible to occur
either explicitly or spontaneously in the Higgs sector of the MSSM at the tree level.
If the $\mu$ parameter and the soft supersymmetry breaking parameters may possess complex phases,
the redefinition of Higgs fields can always eliminate them.
A global phase rotation can further eliminate any complex phases in the vacuum expectation
values of two Higgs doublets.
At the one-loop level, it has been studied that explicit CP violation is possible in the MSSM,
but spontaneous CP violation scenario is difficult to be realized in the MSSM,
because the mass of the lightest neutral Higgs boson in the scenario turns out to be very light
to satisfy the LEP2 result for Higgs search [9-11].

Quite recently, a model has been proposed by Dine, Seiberg, and Thomas within
the framework of the effective field theory analysis [12].
They have shown that the corrections from the new physics beyond the MSSM may be described in
terms of higher-dimensional operators below the scale of the new physics.
The simplest version, which we call the Dine-Seiberg-Thomas model (DSTM), has
just two dimension-five operators with the MSSM particle content in its Higgs sector [13-16].

We have been attracted by the possibility that the DSTM may also accommodate CP violation in its Higgs sector.
We have found that the DSTM may indeed allow spontaneous CP violation
at the one-loop level, for wide ranges of parameters values,
where top quark and scalar top quark loops are taken into account [17].
In the spontaneous CP violation scenario, the upper bound on the radiatively corrected mass of
the lightest neutral Higgs boson in the DSTM is calculated to be 87 GeV.
This value seems to be too small, but it does not contradict with the LEP2 data,
as the relevant coupling coefficients are also small.

This article is the second part of our study on the DSTM within the context of CP violation.
In this article, we study the possibility of explicit CP violation in the DSTM.
In the DSTM, there are three possible sources of explicit CP violation, namely, the
complex phases in the coefficients of two dimension-five operators and in the parameters of
the MSSM sector.

We find that explicit CP violation scenario in the DSTM at the tree level is practically unacceptable,
because the mass of the lightest neutral Higgs boson in the scenario is inconsistent with the LEP2 data.
However, at the one-loop level, the radiative corrections from the top and scalar top quark loops
allow the DSTM to accommodate explicit CP violation for a wide region in the parameter space,
consistent with the LEP2 experimental constraint.

\section{Explicit CP violation at tree level}

The Higgs potential at the tree level may be written as
\begin{eqnarray}
V^0 & = & m_u^2 |H_u|^2 + m_d^2 |H_d|^2 - \bigg (m_{ud}^2 H_u H_d + {\rm H.c.} \bigg )   \cr
&  &\mbox{} + {1 \over 8} ({g'}^2 + g^2) \bigg ( |H_u|^2 - |H_d|^2  \bigg )^2   \cr
&  &\mbox{} + \bigg [ 2 \epsilon_1 ( |H_d|^2 + |H_d|^2 ) H_u H_d    \cr
& &\mbox{} + \epsilon_2 (H_u H_d)^2 + {\rm H.c.} \bigg ]  \  ,
\end{eqnarray}
where $H_d^T = (H_d^0, H_d^-)$ and $H_u^T = (H_u^+, H_u^0)$ are two Higgs doublets,
$g'$ and $g$ are respectively the gauge coupling coefficients for $U(1)$ and $SU(2)$,
$m_u$, $m_d$, and $m_{ud}$ are the mass parameters,
and $\epsilon_{1,2}$ are the coefficients for two effective dimension-five operators in the DSTM.
These new coefficients represent the higher-dimensional interactions as quartic Higgs couplings.

There are several sources of complex phases in the tree-level Higgs potential:
$\epsilon_1$ and $\epsilon_2$ may have complex phases $\varphi_1$ and $\varphi_2$, respectively;
$m_{ud}$ may also be generally complex whereas $m_u$ and $m_d$ can be made real.
However, the phase of $m_{ud}$ may be removed by redefining the phases of the two Higgs doublets.
After redefinition of their phases, the Higgs doublets may be written as
\begin{eqnarray}
H_d & = & \left(
\begin{array}{c}
v_d + \phi_d + i \psi_d     \\
H_d^-
\end{array} \right) \ ,      \cr
H_u & = & \left(
\begin{array}{c} H_u^+ \\
v_u + \phi_u  + i \psi_u
\end{array}\right)
\end{eqnarray}
where $v_d$ and $v_u$ are the vacuum expectation values of the neutral Higgs fields.
We assume that they are real.
If CP symmetry is conserved in the Higgs sector, $\phi_d$ and $\phi_u$ would be the scalar fields
while $\psi_d$ and $\psi_u$ would be the pseudoscalar fields,
and a linear combination of the two pseudoscalar fields would be the pseudo-Goldstone mode.

There are two tadpole minimum conditions with respect to $\psi_d$ and $\psi_u$.
If CP be conserved in the Higgs sector, the minimum conditions would be trivial.
In our case of explicit CP violation, the minimum conditions are no longer trivial.
However, the two minimum conditions yield a single identical equation.
This equation gives us a realtionship between complex phases at the tree level, in case of explicit CP violation, as
\begin{equation}
	\sin \varphi_1 = - {\epsilon_2 \over \epsilon_1} \cos \beta \sin \beta \sin \varphi_2
\end{equation}
where $\tan\beta = v_u/v_d$ and $\varphi_i$ are the complex phases of $\epsilon_i$.
We may use this relationship to reduce the number of indepnedent complex phases.
Therefore, in the following expressions, $\varphi_1$ is not an independent variable but
simply an expression depending on $\epsilon_i$, $\tan \beta$, and $\varphi_2$.

The matrix elements of $M^0$, the symmetric $3 \times 3$ mass matrix for the neutral Higgs bosons at the tree level, are
given explicitly as
\begin{eqnarray}
M_{11}^0 & = & m_Z^2 \cos^2 \beta + m_{ud}^2 \tan \beta   \cr
& &\mbox{} + 2 \epsilon_1 v^2 ( 1 + 2 \cos 2 \beta ) \tan \beta \cos \varphi_1      \ , \cr
M_{22}^0 & = & m_Z^2 \sin^2 \beta + m_{ud}^2 \cot \beta   \cr
& &\mbox{} + 2 \epsilon_1 v^2 ( 1 - 2 \cos 2 \beta ) \cot \beta \cos \varphi_1   \ , \cr
M_{33}^0 & = & {|m_{ud}|^2 \over \cos \beta \sin \beta}
- {2 \epsilon_1 v^2 \cos \varphi_1 \over \cos \beta \sin \beta} \cr
& &\mbox{} - 4 \epsilon_2 v^2 \cos \varphi_2  \  ,   \cr
M_{12}^0 & = &\mbox{} - m_Z^2 \cos \beta \sin \beta - m_{ud}^2 + 6 \epsilon_1 v^2 \cos \varphi_1  \cr
& &\mbox{} + 2 \epsilon_2 v^2 \sin 2 \beta \cos \varphi_2   \ , \cr
M_{13}^0 & = &\mbox{} - 6 \epsilon_1 v^2 \cos \beta \sin \varphi_1   \cr
& &\mbox{} - {1 \over 2} \epsilon_2 v^2 ( 5 \sin \beta + \sin 3 \beta ) \sin \varphi_2   \ , \cr
M_{23}^0 & = &\mbox{} - 6 \epsilon_1 v^2 \sin \beta \sin \varphi_1   \cr
& &\mbox{} - {1 \over 2} \epsilon_2 v^2 ( 5 \cos \beta - \cos 3 \beta ) \sin \varphi_2   \ ,
\end{eqnarray}
where $v  = \sqrt{v_d^2 + v_u^2} = 175$ GeV and
$m_Z^2 = ({g'}^2+g^2) v^2/2$ is the squared mass of $Z$ boson.

Among them, the CP violation is triggered by  $M_{13}$ and $M_{23}$, which mix the scalar and pseudoscalar Higgs fields.
If $\varphi_1 = \varphi_2 = 0$, the CP symmetry would be conserved, and both $M_{13}$ and $M_{23}$ would be zero.
In this case, $M_{33}^0$ would be the squared mass of the pseudoscalar Higgs boson of the DSTM,
and the two eigenvalues of the upper-left $2\times 2$ submatrix of $M^0$ would be
the squared masses of two scalar Higgs bosons.

Note that, if $\epsilon_1 = \epsilon_2 = 0$, only the first term of $M_{33}^0$ remains, which is identical to
the squared mass of the pseudoscalar Higgs boson at the tree level in the CP-conserving MSSM,
$m^2_{A^0} = |m_{ud}|^2 /\cos \beta \sin \beta$.

In the DSTM, they are not zero in general, and therefore the pseudoscalar Higgs boson in the DSTM has additional contributions
from the dimension-five operators, the second and the third terms of $M_{33}^0$,
These higher-dimensional contributions do not vanish even if $\varphi_1 = \varphi_2 = 0$,
that is, in the CP-conserving limit.
It should also be noticed that
$\varphi_1$ and $\varphi_2$ are responsible for the scalar-pseudoscalar mixings at the tree level.

By diagonalizing $M^0$, three eigenvalues are calculated.
They are sorted in increasing order to obtain three squared masses for the neutral Higgs bosons in the DSTM
in the explicit CP violation scenario.
The upper bound on the mass of the lightest neutral Higgs boson in the DSTM is obtained as
\begin{eqnarray}
m_{h_1}^2 & \le & m_Z^2 \cos^2  2 \beta + 8 \epsilon_1 v^2 \sin  2 \beta \cos \varphi_1 \cr
& &\mbox{} + 2 \epsilon_2 v^2 \sin^2  2 \beta \cos \varphi_2 \ .
\end{eqnarray}
Note that, if $\epsilon_1$ and $\epsilon_2$ are positive, the mass of the lightest neutral Higgs boson
increases.
On the other hand, in the CP-conserving DSTM, the mass of the pseudoscalar Higgs boson decreases if
$\epsilon_1$ and $\epsilon_2$ are positive.

For the numerical analysis, we choose $m_{A^0}$, the physical tree-level mass of the pseudoscalar Higgs boson
in CP-conserving MSSM, instead of $m_{ud}$ as a free parameter.
The ranges for the free parameters are set as follows:
$2 < \tan \beta < 30$, $|\epsilon_1| < 0.025$, $|\epsilon_2| < 0.025$,
$|\varphi_2| < \pi/2$, and $0 < m_{A^0} < 1000$ GeV.
We first calculate the tree-level masses of the three neutral Higgs bosons in the DSTM.
We find that $m_{h_1}$ may be as large as about 98 GeV within our parameter space.
The allowed ranges for the masses of the other neutral Higgs bosons are $15 < m_{h_2} < 1017$ GeV
and $91 < m_{h_3} < 1017$ GeV.

Then, we calculate the coupling coefficients of the three neutral Higgs bosons to a pair of $Z$ bosons,
and normalize them with respect to the corresponding quantity in the SM.
The normalized coupling coefficients are given by
\begin{equation}
G_{ZZh_i} = \cos \beta O_{1i} + \sin O_{2i} \ ,
\end{equation}
where $O_{ij}$ are the elements of the orthogonal matrix that diagonalizes $M^0$.

In terms of these masses and the normalized $ZZh_i$ coupling coefficients, we examine
the discovery limit of any one of the three neutral Higgs bosons in the DSTM with explicit CP violation,
by comparing them with the experimental results for the Higgs search at LEP2
to all the Higgs couplings to a $Z$ boson pair [18].

We find that at least one of the three neutral Higgs bosons has a very strong coupling
to a pair of $Z$ bosons,
if its mass is less than 114.5 GeV.
This implies that at least one of the three neutral Higgs bosons in the DSTM is
above the discovery limit of LEP2.
The negative results from LEP2 thus exclude the possibility of explicit CP violation
in the DSTM at the tree level.

\section{Explicit CP violation at one-loop level}

Now, we turn our attention to the possibility of explicit CP violation in the DSTM at the one-loop level.
The procedure is quite similar to the tree-level analysis.

On the tree-level mass matrix for the scalar top quarks,
$\mu$ and $A_t$ may also be complex in general.
The complexity of $\mu$ and $A_t$ result in the expression for the scalar top quark masses after
electroweak symmetry breaking as
\begin{equation}
m_{{\tilde t}_1, {\tilde t}_2}^2 =  {(m_Q^2 + m_T^2) \over 2} + m_t^2
+ {m_Z^2 \over 4} \cos 2 \beta \mp \sqrt{X_t} \ ,
\end{equation}
with
\begin{eqnarray}
X_t & = & \bigg [ {m_Q^2 - m_T^2 \over 2} + \bigg ({2 \over 3} m_W^2 - {5 m_Z^2 \over 12}  \bigg ) \cos 2 \beta \bigg ]^2 \cr
& &\mbox{} + m_t^2 \bigg ( A_t^2 + {\mu^2 \over \tan^2 \beta} - {2 \mu A_t \cos \varphi \over \tan \beta} \bigg ) \ ,
\end{eqnarray}
where
$m_W^2 = g^2  v^2 /2$ is the squared mass of $W$ boson,
the top quark mass is given by $m_t = h_t v_u $, where $h_t$ is the Yukawa coupling for the top quark,
and $\varphi$ is the overall phase of $\mu$ and $A_t$.
Note that $X_t$ represents the mixing between the left-handed and the right-handed scalar top quarks.
We do not neglect the $D$-term contributions for the scalar top quark masses,
as the weak-gauge couplings $g'$ and $g$ are included in the above formulae.

The one-loop correction to the Higgs potential is calculated from the effective potential method as [19]
\begin{eqnarray}
V^1 & = & \sum_{i = 1}^2 {3 {\cal M}_{{\tilde t}_i}^4 \over 32 \pi^2}
    \bigg [ \log {{\cal M}_{{\tilde t}_i}^2 \over \Lambda^2} - {3 \over 2} \bigg ]   \cr
    & &\mbox{} - {3 {\cal M}_t^4 \over 16 \pi^2}
    \bigg [ \log {{\cal M}_t^2 \over \Lambda^2} - {3 \over 2} \bigg ]                 \ ,
\end{eqnarray}
where
$\Lambda$ is the renormalization scale in the modified minimal subtraction scheme,
${\cal M}_{{\tilde t}_i}$ $(i=1,2)$ are the field-dependant masses for  the scalar top quarks,
and ${\cal M}_t$ is the field-dependent top quark mass.
The total Higgs potential at the one-loop level is thus given by $V = V^0 + V^1$.

The mass matrix for the three neutral Higgs bosons at the one-loop level receive contributions from $V^1$.
Denoting the one-loop contribution as $M^1$, the full mass matrix is expressed as
\begin{equation}
    M = M^0 + M^1 \ .
\end{equation}
The expressions for $M^1_{ij}$ ($i,j = 1,2,3$) are identical to the expressions in Ref. [17],
where the one-loop contribution is calculated in case of spontaneous CP violation.

As mentioned before, two complex phases, $\varphi_1$ and $\varphi_2$, are present in $M^0$.
Now, in $M^1$, there is the third one, $\varphi$.
Thus, these three complex phases may trigger explicit CP violation in the DSTM at the one-loop level.
However, as in the tree-level case, the tadpole minimum equations at the one-loop level with respect to $\psi_u$ and $\psi_d$
reduce the number of independent complex phases.
Actually, only one of equations is non-trivial in the explicit CP violation scenario.
The equation between complex phases may be expressed as
\begin{eqnarray}
\sin \varphi_1 & = &\mbox{} - {\epsilon_2 \over \epsilon_1} \cos \beta \sin \beta \sin \varphi_2   \cr
& & \mbox{} - {3 m_t^2 \mu A_t \sin \varphi \over 32 \pi^2 v^4 \sin^2 \beta \epsilon_1}
f (m_{{\tilde t}_1}^2,  \ m_{{\tilde t}_2}^2)  \   ,
\end{eqnarray}
where the second term is the correction at the one-loop level, with a dimensionless function defined as
\begin{equation}
 f(m_x^2, \ m_y^2) = {1 \over (m_y^2 - m_x^2)} \left[  m_x^2 \log {m_x^2 \over \Lambda^2} - m_y^2
\log {m_y^2 \over \Lambda^2} \right] + 1 \ .
\end{equation}
Using this equation, we may eliminate $\varphi_1$, and the remaining $\varphi_2$ and $\varphi$ become
responsible for the CP mixing between scalar and pseudoscalar Higgs fields at the one-loop level.

At the one-loop level, the upper bound on the mass of the lightest neutral Higgs boson in the explicit CP violation scenario is modified as
\begin{eqnarray}
m_{h_1}^2 & \le & m_Z^2 \cos^2 (2 \beta) + 8 \epsilon_1 v^2 \sin (2 \beta) \cos (\varphi_1) \cr
& &\mbox{} + 2 \epsilon_2 v^2 \sin^2 (2 \beta) \cos (\varphi_2) + \Delta m_{h_1}^2 \ ,
\end{eqnarray}
where $\Delta m_{h_1}$ comes from the radiative corrections due to the loops of top and scalar top quarks.
Its expressions is complicated.

We take the mass of top quark as 171 GeV
and assume that the mass of top quark is smaller than the masses of the scalar top quarks.
We calculate the masses of the three neutral Higgs bosons of the DSTM at the one-loop level.
At the one-loop level, the allowed ranges for the relevant parameters are as follows:
$2 < \tan \beta < 30$, $|\epsilon_1| < 0.025$, $|\epsilon_2| < 0.025$, $|\varphi_2| < \pi/2$,
$|\varphi| < \pi/2$, $0 < m_{A^0} < 1000$ GeV, $100 < |\mu| < 500$ GeV, $|A_t| < 1000$ GeV,
and $50 < m_Q = m_T < $  500 GeV.
Here, the experimental data on the chargino system are used to set the lower bound on the absolute value of $\mu$.
The upper bound on the mass of $h_1$ is about 137 GeV.
The masses of the heavier neutral Higgs bosons are calculated to be $16 < m_{h_2} < 1073$ GeV
and $112 < m_{h_3} < 1074$ GeV.
We show 50,000 points of ($\tan\beta$, $m_{h_1}$) in Fig. 1, which are obtained
by randomly choosing the parameter values in the respectively allowed ranges.
One may notice that most of the points in Fig.1 are above the line $m_{h_1} = 114.5$ GeV, the LEP2 lower
bound on the mass of the scalar boson.
These points are acceptable, since the negative results for the Higgs search at LEP2 do not exclude these points.
However, some of the points are below the LEP2 line.
Thus, these points are certainly within the discovery limit of LEP2.
We should confirm that the points below the LEP2 line are indeed consistent with the LEP constraints.

For each set of parameter values that yield $m_{h_i} < 114.5$ GeV,
we calculate the normalized coupling strengths of the three
neutral Higgs bosons to a pair of $Z$ bosons at the one-loop level.
These normalized coupling strengths would indeed tell whether the neutral Higgs bosons of the DSTM might
have escaped LEP2 or not.

In particular, we examine the normalized coupling coefficients $G_{ZZh_1}$ of the lightest Higgs boson.
By randomly varying the parameter values within the same allowed ranges as in Fig. 1,
we calculate $G^2_{ZZh_1}$  and $m_{h_1}$.
In Fig. 2, we show 50,000 points of ($m_{h_1}$,$G^2_{ZZh_1}$) of the lightest Higgs boson at the one-loop level.
The general tendency of the distribution in Fig. 2 is that the normalized coupling coefficient becomes
larger as the lightest Higgs boson becomes heavier.
As the mass of the lightest neutral Higgs boson approaches 114.5 GeV, the LEP2 lower bound on the mass of the SM Higgs boson,
the normalized coupling strength grows to 1, that is, it becomes nearly equal to the SM coupling coefficient.
This implies that for the parameter values which yield $m_{h_1} \sim 114.5$ GeV and $G_{ZZh_1} \sim 1.0$,
the lightest Higgs boson behaves more or less like the SM Higgs boson and
the contributions of the heavier Higgs bosons become negligible with respect to decays into a pair of $Z$ bosons.

The result shown in Fig.2 suggests that the neutral Higgs boson of the DSTM with $m_{h_1} < 114.5$ GeV
could not be detected at LEP2.
In other words, the explicit CP violation scenario in the DSTM at the one-loop level is consistent with the LEP2 constraints,
for the parameter space examined above.
Therefore, the possibility of explicit CP violation in the DSTM at the one-loop level
is allowed by the present Higgs phenomenology.

\begin{figure}[t!]
\includegraphics[width=3in]{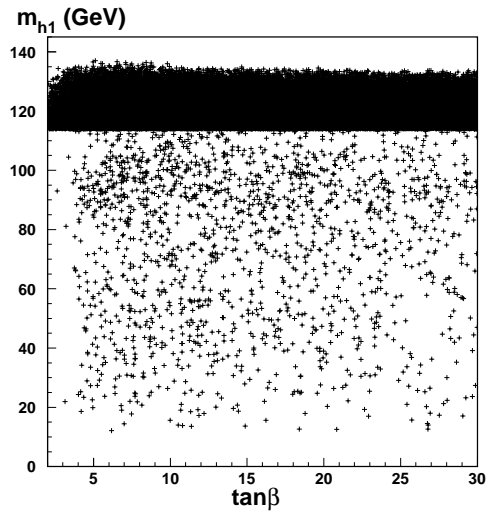}
\caption{The distribution of 50,000 points of ($\tan\beta$, $m_{h_1}$), at the one-loop level.
The allowed ranges of the parameter values are:
$|\epsilon_1| < 0.025$, $|\epsilon_2| < 0.025$, $|\varphi_2| < \pi/2$,
$|\varphi| < \pi/2$, $0 < m_{A^0} < 1000$ GeV, $100 < |\mu| < 500$ GeV, $|A_t| < 1000$ GeV,
and $50 < m_Q = m_T < 500$ GeV.
Note that the points are evenly distributed with respect to $\tan\beta$, showing no dependence of $m_{h_1}$ on $\tan\beta$.
This feature of the DSTM is different from the CP-conserving MSSM, where the maximum of $m_{h_1}$ occurs for large $\tan \beta$.
}
\end{figure}
\begin{figure}[t!]
\includegraphics[width=3in]{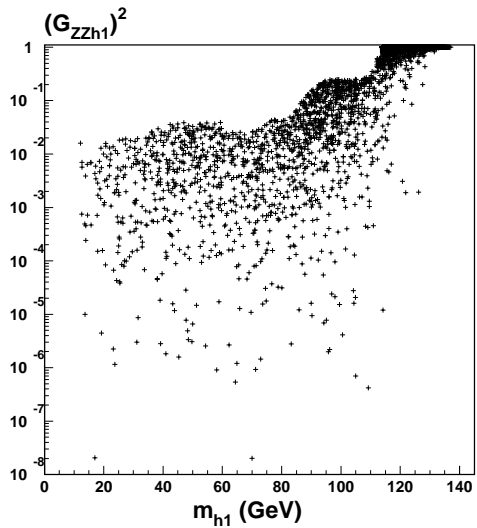}
\caption{The distribution of 50,000 points of ($m_{h_1}$, $G^2_{ZZh_1}$),
the square of normalized coupling strength of
the lightest Higgs boson of the DSTM versus its mass, at the one-loop level.
The allowed ranges of the parameter values are the same as in Fig. 1.
}
\end{figure}
%
\section{Conclusions}

We examine the possibility of explicit CP violation, for a reasonable parameter space of the DSTM.
At the tree level, the upper bound on the lightest neutral Higgs boson is about 98 GeV.
Thus, they should have been discovered at LEP2 unless their couplings to a pair of $Z$ bosons are
very small compared with the SM coupling coefficients.
We explore randomly the parameter space to find that, for any set of parameter values, at least one of
the three neutral Higgs bosons has its mass below 114.5 GeV and at the same time its
coupling coefficients to a pair of  $Z$ bosons very large.
This implies that any set of parameter values in the parameter space is inconsistent with the LEP2
experimental data.
In other words, the explicit CP violation in the DSTM at the tree level is not possible.

The situation is clearly improved at the one-loop level.
For most of parameter values, the mass of the lightest neutral Higgs boson is calculated to be larger than 114.5 GeV.
For those parameter values that yield the mass of the lightest neutral Higgs boson below 114.5 GeV,
the coupling coefficient to a pair of  $Z$ bosons are calculated to be very small,
implying that these points are allowed by the LEP2 constraints.
Hence, the explicit CP violation scenario in the DSTM at the one-loop level is allowed by the LEP2 data.
In conclusion, the DSTM may accommodate explicit CP violation at the one-loop level.

\section{Acknowledgments}
We thank Kihyeon Cho at KISTI for the collaboration.
This research was supported by Basic Science Research Program
through the National Research Foundation of Korea (NRF) funded
by the Ministry of Education, Science and Technology
(2009-0086961, 2009-0070667).


\end{document}